# A Federated Semi-Supervised Learning Approach for Network Traffic Classification

Yao Peng, Meirong He, Yu Wang*

Institute of Artificial Intelligence and Blockchain, Guangzhou University, China

*yuwang@gzhu.edu.cn

**Abstract.** Network traffic classification, a task to classify network traffic and identify its type, is the most fundamental step to improve network services and manage modern networks. Classical machine learning and deep learning method have developed well in the field of network traffic classification. However, there are still two major challenges. One is how to protect the privacy of users' traffic data, and the other is that it is difficult to obtain labeled data in reality. In this paper, we propose a novel approach using federated semi-supervised learning for network traffic classification. In our approach, the federated servers and several clients work together to train a global classification model. Among them, unlabeled data is used on the client, and labeled data is used on the server. Moreover, we use two traffic subflow sampling methods: simple sampling and incremental sampling for data preprocessing. The experimental results in the QUIC dataset show that the accuracy of our federated semi-supervised approach can reach 91.08% and 97.81% when using the simple sampling method and incremental sampling method respectively. The experimental results also show that the accuracy gap between our method and the centralized training method is minimal, and it can effectively protect users' privacy and does not require a large amount of labeled data.

## 1 Introduction

Traffic classification [1] plays a vital role both in the management of the modern network and the and security of the Internet. The goal of the traffic classification task is to classify Internet traffic into predefined categories, such as protocol category (UDP, TCP, etc.), application type (Netflix, Youtube, Facebook, etc.), or abnormal detection(Normal or Abnormal). Not only can it be used by Internet Service Providers (ISP) for quality of service (QoS) and troubleshooting tasks, it can also be used in intrusion detection systems to protect users' devices from hackers[2]. Therefore, the task of traffic classification has always been the focus of network security research.

In the early development of the Internet, the identification of network traffic was a simple task, and methods based on ports and deep packet inspection (DPI) were usually used. However, with the increasing number of new applications, many applications no longer have fixed port numbers that can be queried, and popular P2P applications generally adopt random port strategies. Therefore, the accuracy of methods based on port identification continues to decline and gradually has been phased out. Although DPI based methods have higher accuracy, it has disadvantages such as higher computational complexity and the inability to process encrypted traffic.

---

* represents corresponding author

To overcome the shortcomings of port identification and DPI, researchers try to utilize machine learning methods to perform network traffic classification tasks. Among them, more common is based on statistical machine learning methods. The core is to extract statistics features from flow generated by different types of applications, then select some kind of machine learning model (such as Support Vector Machines (SVM) [3, 4], K Nearest Neighbor (KNN)[5, 6], K-means[7], Decision Tree[8], etc) to train. The advantage of this method is that the computational complexity is low and it can be applied to encrypted traffic. But its disadvantages are also obvious that it is severely dependent on manual designed features.

The emergence of deep learning[9] has greatly improved the feature dependence problem of traditional machine learning. In recent years, there have been many pieces of research on applying deep learning to network traffic classification tasks, such as[10-13], etc. These existing studies have shown that deep learning performs well on public network traffic data sets, however, deep learning still faces some difficulties in practical applications:

- The issue of privacy and security: In the field of network traffic classification, the use of deep learning technology often faces the problem of data privacy leakage because the traffic data collected from user devices often contain private information about network behaviors. Generally, users do not want this information to be disclosed.
- The issue of data island: The success of deep learning technology lies in a large amount of training data. However, as mentioned above, users' network traffic data often contain private information that laws or regulations do not allow to disclose. Therefore, each company or institution is like an independent island, storing and defining data internally, which resulting in obtain an overfitting model when training with these homogenous data, and the accuracy of the model will be greatly reduced in practical applications.
- The issue of lacking labeled data: Deep learning technology is still mainly based on supervised learning, and it needs to collect a large amount of labeled data to train the model. However, in reality, most of the collected data is untagged, and due to the complexity of computer network domain knowledge, a large number of professionals are required to label data, which will consume huge labor and time costs.

To tackle the problems mentioned above that are encountered in the practical application of deep learning, we propose a novel approach using federated semi-supervised learning to conduct the network traffic classification task in this paper, which provides an attractive trade-off between practicality and privacy protection. Our contributions are as follows:

- Our approach is based on the federated learning environment, an exceptional distributed machine learning and design to protect individual data. Therefore, our method allows multiple parties to jointly learn a traffic classification model without disclosing and sharing their local user data sets. Our method not only subtly resolves the problem of exposing user privacy data but also solves the problem of data islands in the traffic field.
- We design different classification models based on convolutional neural networks to achieve semi-supervised learning in the above-mentioned federated environment. These models combine unsupervised learning and supervised learning. Thus, our method can use a large amount of unlabeled data and a small amount of labeled data simultaneously while training a model, which can effectively reduce the high cost of labeling data in real network traffic classification tasks.
- By combining the advantages of federal learning and semi-supervised learning, our approach is more practical than the previous centralized deep learning method.

## 2 Related Work

Deep learning has achieved great success in image classification, speech recognition, and other research directions for the past few years. Affected by this, more and more research based on deep learning has appeared in network traffic classification. Wang et al.[11] proposed a one-dimensional Convolutional Neural Network (1D-CNN) using a one-dimensional vector to represent each flow or session as input. The model was evaluated on 12 types of encrypted application datasets and showed a significant improvement over the C4.5 method with time series and statistical features. Chen et al.[12] proposed 2D-CNN with two convolutional layers, two pooling layers, and three fully connected layers. This method uses reproducing kernel Hilbert space (RKHS) embedding and converts the flow time series data into a two-dimensional image. The result of this method is superior to traditional machine learning methods in terms of protocol and application classification. In [13], Lopez-Martin et al. adopted the framework composed of CNN and Recursive Neural Network (RNN), providing the best detection results for the RedIRIS dataset. These studies show that deep learning outperforms the traditional machine learning methods on the network traffic classification task, but it can also be seen that they are all supported by a large number of labeled datasets which is challenging to obtain in reality.

In order to address the paucity of labels for network traffic data, many works based on semi-supervised deep learning have been proposed. Iliyasu et al.[14] demonstrated a method based on deep convolution generated adversative networks (DCGANs) to improve the performance of a classifier trained on a few labeled samples. Rezaei and Liu et al.[15] adopted the transfer learning method based on deep neural networks to classify network traffic, which avoids numerous labeled datasets. The accuracy of their method is almost the same as the fully supervised method of labeling large datasets. [16] and [17] employ the automatic encoder, which is a common technique in semi-supervised learning. In [16], the author using stacked sparse autoencoder (SSAE) accompanied by de-noising and dropout techniques to integrate feature extraction and classification into a model. The obtained results demonstrate a better performance than traditional models while keeping the whole procedure automated. In [17], the author proposed a variational autoencoder model for anomaly detection. The model outperforms the other semi-supervised learning models with a 5-10 percent increase in evaluation metrics.

Later, federated learning for privacy protection has also been applied to the task of network traffic classification. Hyunsu et al. [18] proposed a federated-learning traffic classification protocol (FLIC). One of the most prominent traits of this protocol is that new applications can be classified in real-time without compromising privacy. In [19], the author proves that the federated model is also comparable to the centralized training approach.

Recently, a topic worthy of attention is the federal semi-supervised learning method proposed in [20]. On the one hand, this technology can ensure user privacy and effectively utilize more training data. On the other hand, semi-supervised learning can alleviate the dependence on labeled data in deep learning. Although the federal semi-supervised technology has just started, there has been work using the federal semi-supervised learning technology to solve the problem and achieved good results, such as [20]. However, there has not yet been work on federated semi-supervised learning technology in network traffic classification, which is novel and conducive to practical applications.

## 3 Problem Definition

In this section, we start with a brief introduction to federated learning and semi-supervised learning. Then, we explain the overview of federated semi-supervised learning architecture.

### 3.1 Federated Learning

Federated learning [21] is primarily about obtaining a common global model through communication between a central server and a local client. The design goal of federated learning is to carry out efficient machine learning among multiple participants or computing nodes under the premise of ensuring information security during big data exchange, protecting the privacy of terminal data and personal data, and ensuring legal compliance. The federated learning process can be broken down into four steps as shown in Fig. 1: (1) Participants use the local data training model and the encryption gradient is uploaded to server A. (2) Server A updates model parameters by aggregating the gradient of each user. (3) Server A returns the updated model to the participants. (4) Each participant updates their model.

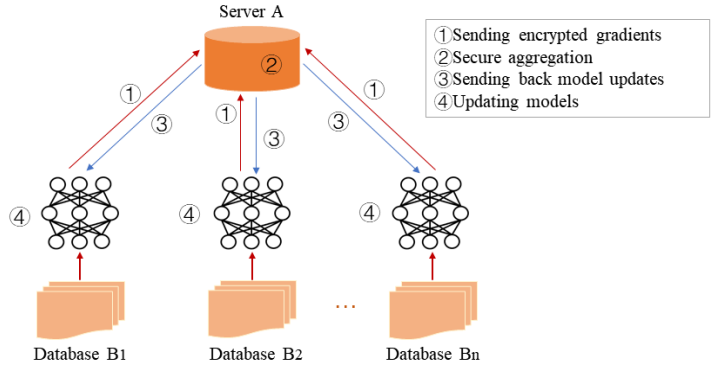

**Fig. 1.** The process of federated learning

The details of the federated learning process are as followed. Let $M_G$ be the global model obtained on the server and $M_L = \{M_{L_k}\}_{k=1}^{k}$ be the local model for $k$ clients. $D_s = \{x_i, y_i\}_{i=1}^{N}$ is sub-dataset used for training by the local client, which is composed of the number of $N$ instance with training instance $x_i$ and the corresponding label $y_i$. Before the local client trains $M_{L_k}$, $D_s$ needs to be assigned to each local client. At each communication round $t + 1$ between server and clients, $M_G$ needs to aggregate the learning weights $w$ in $M_{L_k}$ of $k$ local clients to get a new $w_{G_{t+1}} \leftarrow w_{G_t}$ until the end of the last round of communication.

### 3.2 Semi-supervised

Semi-supervised Learning (SSL) is a Learning method that combines unsupervised Learning with Supervised Learning, which uses large amounts of unlabeled data and small amounts of labeled data simultaneously to train a model (Fig.2). Under the guidance of a small number of labeled data, semi-supervised learning can make full use of a large number of unlabeled data to improve learning performance and avoid the waste of data resources. At the same time, it solves the problem that the generalization ability of supervised learning is not strong when there are few labeled data, and the unsupervised learning method is not accurate when there is no labeled data guidance.

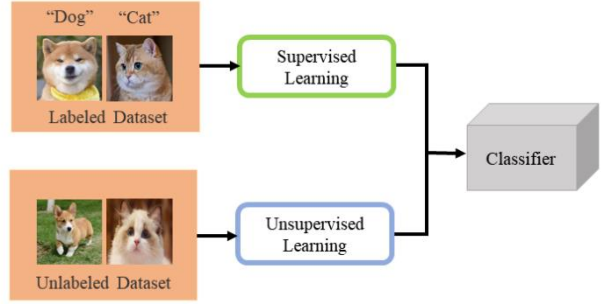

**Fig. 2.** Semi-supervised learning

The normal form datasets $D = \{d_s, d_u\}$ contains labeled data $d_s = \{x_i, y_i\}_{i=1}^{S}$ and unlabeled dataset $d_u = \{x_i\}_{i=1}^{U}$, in general $|d_s|''|d_u|$. $d_s$ and $d_u$ respectively perform supervised learning and unsupervised learning, input $x_i$ into the deep network to predict $\hat{y}_i$, and finally, obtain the loss function $loss_s$ based on supervision and the loss function $loss_u$ based on unsupervised. Our ultimate goal is to minimize the loss function $loss_f = loss_s + loss_u$, to obtain a network model with strong learning ability.

### 3.3 Federated Semi-supervise Learning

The application of semi-supervised learning under the federated framework is called federated semi-supervised learning (FSSL). Under the general FSSL architecture, there are a large number of unlabeled data $d_u = \{x_i\}_{i=1}^{U}$ and a group of local model $M_L = \{M_{L_k}\}_{k=1}^{k}$ at local clients, and there are a small number of labeled data $d_s = \{x_i, y_i\}_{i=1}^{S}$ and a global model $M_G$ on sever. The application scenarios we designed will be described in detail in the following chapters.

## 4 Methodology

In this section, we introduce the architecture of the traffic classification approach based on FSSL and introduce the three stages of it. They are the data preprocessing stage, the federated pretraining stage, and the central server retraining stage. In addition, the rules of simple sampling and incremental sampling used in the data preprocessing stage and the model used in the approach are also described in detail.

### 4.1 FSSL Traffic Classification Approach Architecture

By taking advantage of federated learning and semi-supervised learning, a federated semi-supervised traffic classification approach architecture has been proposed, as illustrated in Fig. 3. As shown in the figure, the approach is based on a federated environment, that is, a federated central server (hereinafter referred to as the central server) and several clients, using semi-supervised learning to train the network traffic classifier. It contains three stages, including the data preprocessing stage, the federated pretraining stage, and the central server retraining stage.

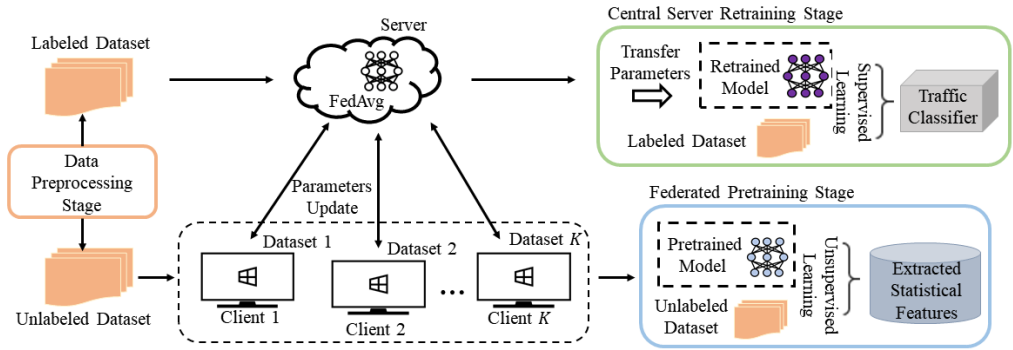

**Fig. 3.** Three stages of the federal semi-supervised traffic classification approach. The yellow box shows the data preprocessing stage, the green box shows the pretraining stage, and the blue box shows the central server retraining stage.

In order to reduce computational complexity and compress storage data space, we use simple sampling and incremental sampling methods to process the unlabeled network traffic dataset on each client and the labeled network traffic dataset on the central server in the data preprocessing stage. The specific method is to extract multiple sets of sampled data packets for each network flow according to the rules of simple sampling and incremental sampling introduced in 4.2. A set of sampled data packets is composed of a subflow, then each sample of the network traffic dataset will be input into the training model in the form of multiple subflows. In addition, we need to extract 24 features (such as average packet length, average inter-arrival time, etc.) of each flow of the unlabeled dataset on the client as the target of the regression function of the federated pretrained model.

In the federated pretraining stage, we select several clients. In each communication round between the client and the central server, the selected client uses its local unlabeled data to perform unsupervised training and then sends the trained model weights to the central server. The central server uses an aggregation algorithm to aggregate the weights of each client to update the server-side global pretrained model. Until the loss of the global model stabilizes, the communication between the client and the server ends. In 4.3, we detailed the pretrained model used in this stage.

In the central server retraining stage, the central server uses a small portion of the labeled dataset it owns to retrain the global pertained model and generate the final network traffic, classification model. After retraining, each client can download the final classification model to its local to perform network traffic classification tasks. In 4.3, we detailed the retrained model used in this stage.

To more precisely describe the methods used in our scheme, we define the following variables: $K$ represents the number of clients that can participate. $C$ is a decimal in the range of 0~1. It represents the participation rate of the client. So the set of clients participating in federated training in each round can be expressed as $Set\_clients = \{S_k\}_{k=1}^{C \cdot K}$, where $S_k$ represents the $k$th client and $k \in \{1,2, \dots, C \cdot K\}$. $D_u = \{x_i, t_i\}_{i=1}^{n}$ represents the collection of the unlabeled dataset on all clients after preprocessing, where $x_i$ represents the $i$th unlabeled sample, and $t_i$ represents the 24 features extracted by $x_i$ which is the target for unsupervised training. $i \in \{1,2, \dots, n\}$, and $n$ is the number of unlabeled samples. $D_{uS_k}$ represents the unlabeled dataset owned on the $k$th client, and $D_{uS_k} \in D_u$. $D_s = \{x_j, y_j\}_{j=1}^{m}$ represents the labeled dataset on the central server after preprocessing, where $x_j$ represents the $j$th labeled sample, $y_j$ represents the ground truth label of $x_j$, $j \in \{1,2, \dots, m\}$ and $m$ is the number of labeled samples. Generally, n is much larger than m. $r \in \{1,2, \dots, R\}$ represents the communication round in federated training.

The network traffic classification based on FSSL used in our approach can be described as follows: Firstly, $K$ clients obtain local unlabeled network data and features of each flow by data preprocessing. For each client, its local dataset is part of $D_u$. Similarly, the central server obtains a labeled dataset $D_s$. Then the central server respectively initializes the global pretrained model $M_{P_g}$ and its weight $\theta_{P_g}$ and the retrained model $M_R$ and its weight $\theta_R$. When the central server communicates with clients for the $r$th time, the server selects several clients $Set\_clients = \{S_k\}_{k=1}^{C \cdot K}$ according to the formula $max(C \cdot K, 1)$ and send the current weight $\theta_{P_g}^r$ of model $M_{P_g}$ to each client. Where $max(C \cdot K, 1)$ means that if $C \cdot K$ is a number less than 1, the server selects at least one client for communication. The selected client $S_k$ receives the weight $\theta_{P_g}^r$ sent by the central service and then performs unsupervised training with its local dataset $D_{uS_k}$ updating the local pre-trained model $M_{P_{S_k}} \leftarrow M_{P_g}$. During this step, the model $M_{P_{S_k}}$ can learn the characteristics of local network traffic data. After all selected clients have completed training, they upload the learned weight $\theta^r_{P_{S_k}}$ of the model $M_{P_{S_k}}$ to the central server. The central server uses the average algorithm FedAvg[22] to obtain the aggregated weight $\theta^{r+1}_{P_g}$ by calculating all weight sent by each client and uses it to update the model $M_{P_g}$. After that, the $r$th round ends, and the central server and clients then execute the next communication round, iterating until the model $M_{P_g}$ on the server converges. After the federal communication process is over, the central server copies the final learning weight $\theta'_{P_g}$ of the model $M_{P_g}$ to the corresponding network layer of the retrained model $M_R$ and then uses the labeled dataset $D_s$ for supervised learning until $M_R$ converges to $M_C$. Clients that need to perform classification tasks could download the model $M_C$ from the central server. In addition, both model $M_{P_g}$ and model $M_R$ use Adam[23] for optimization, and their loss functions are MSELoss and CrossEntropyLoss, respectively. More details of the procedures for our method are shown in Algorithm 1.

---

**Algorithm 1** Federated Semi-supervised.
**Server executes:**
  Initialize $\theta_{P_g}$, $\theta_R$
  **for** each communication round $r \in [1, 2, ...R]$ **do**
    $m \leftarrow max(C \cdot K, 1)$
    $Set\_clients \leftarrow$ The central server selects $m$ clients
    **for** each client $S_k \in Set\_clients$ **do**
      $\theta^r_{P_{S_k}} \leftarrow ClientUpdate(\theta^r_{P_g}, D_{uS_k})$
    **end for**
    $\theta'_{P_g} \leftarrow \theta^{r+1}_{P_g} \leftarrow \sum_{S_k=1}^{Set\_clients} \frac{n_{S_k}}{n} \theta^r_{P_{S_k}}$     $\triangleright FedAvg$
  **end for**
  $copy(\theta_R, \theta'_{P_g})$
  $\theta_C \leftarrow Adam(\theta_R, D_s)$

**ClientUpdate**($\theta^r_{P_g}, D_{uS_k}$):
  $\theta^r_{P_{S_k}} \leftarrow Adam(\theta^r_{P_g}, D_{uS_k})$
  return $\theta^r_{P_{S_k}}$ to Server

---

### 4.2 Sampling Method

The sampling method in the data preprocessing stage is mainly aimed at obtaining a valid representation of the network flow, reducing computational complexity, and alleviating the pressure of massive data storage when training the deep neural network. Specifically, different parts of the flow can be observed by different sampling methods which extracting packets in the network flow. In our work, we adopted simple sampling and incremental sampling methods to process our dataset. The specific settings are as follows:

1) Simple sampling: The packets in each network flow of the dataset are sampled 100 times at an interval $d = 22$. The positions of the packets that start sampling each time are different (increments of 13), and the number of packets sampled each time is 45. Each time of sampling generates a subflow. If a subflow with less than 45 packets, we choose to discard it.

2) Incremental sampling: The packets in each network flow of the dataset are sampled 100 times at an interval $l$. Similarly, the positions of the packets that start sampling each time are different (increments of 13), and the number of packets sampled each time is 45. Each time of sampling generates a subflow. Subflow with less than 45 packets is discarded. The difference from simple sampling is that $l$ is not a fixed value. After each $\alpha$ time, $l$ increases by a multiple of $\beta = 1.6$. So, when sampling in $N$th, $l =$

$l + \frac{N}{\alpha} \cdot \beta$.

In addition, for each sampled packet, we choose the packet length and the relative arrival time as its representation. Network traffic is usually directional, and they are divided into forward and backward, so we multiply the length of the packet by the directional information (that is, + 1 and - 1). When the packet length is positive, it indicates that the packet in the forward flow. Otherwise, it indicates that the packet in the backward flow.

### 4.3 Pretrained Model and Retrained Model

The pretrained model and the retrained model are used in the stage of federated pretraining and the central server retraining respectively. The details of the model structure are presented in Fig. 4. According to the figure, the pretrained model and the retrained model are both based on convolutional neural networks. The difference is that the retrained model has more linear layers and a Softmax layer for multi-classification is added at the end. Both of the models use maximum pooling and choose ReLU as the activation function.

The ability of convolutional neural networks to extract features can help pretrained models learn the characteristics of their respective unlabeled dataset on different clients. After the federation pretrained stage, the learned weight of the global pretrained model in the central server will be transferred to the retrained model, where the learning weight mentioned above is updated by the aggregation of the local pretrained model on each client. Therefore, the retrained model can combine the results of the pretrained model and the labeled dataset to get a better classification effect.

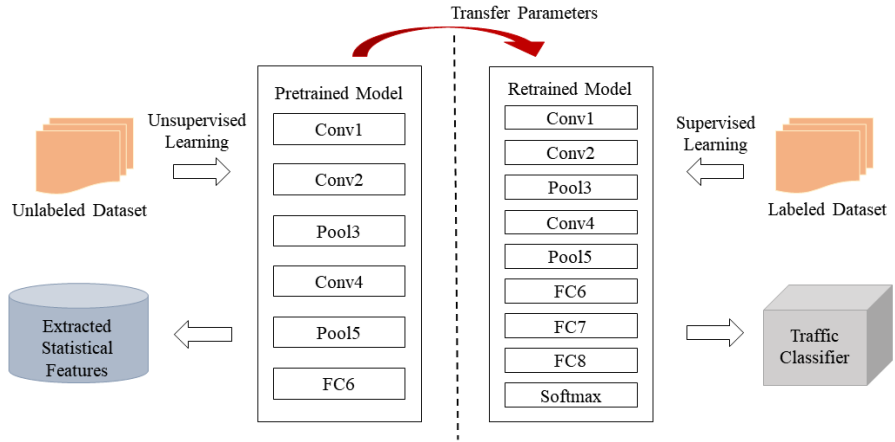

**Fig. 4.** Pretrained model and retrained model

## 5 Experiments

In this section, we evaluated the feasibility of our proposed approach by conducting network traffic classification tasks in the QUIC dataset. In addition, we compare the performance in centralized methods and federated semi-supervised methods.

### 5.1 Performance Metrics

In the experiment we designed, we use the following performance evaluation measures to evaluate the fed-Semi Internet traffic classifier, which are Accuracy, Precision, Recall, and F1-score. They are based on different calculation formulas of cumulative true positives (TP), false positives (FP), true negatives

(TN), and false negatives (FN) of a single class or the entire dataset.

Accuracy is a weighted arithmetic mean of Precision and Inverse Precision (weighted by Bias), as well as weighted arithmetic, mean of Recall, and Inverse Recall (weighted by Prevalence)[24], which is given by:

$$Accuracy = \frac{TP+TN}{TP+TN+FP+FN}. \quad (1)$$

F1-score represents a harmonic mean between precision and Recall. If its value is higher, and the closer to accuracy, the better the classification performance. Precision is the percentage of relevant flows retrieved, while Recall is the percentage of retrieved flows that are relevant.

$$F1 - score = \frac{2}{1/Precision+1/Recall}. \quad (2)$$

$$Recall = \frac{TP}{TP+FN}. \quad (3)$$

$$Precision = \frac{TP}{TP+FP}. \quad (4)$$

## 5.2 QUIC Dataset

We used the QUIC dataset, a public dataset published by [15] in our evaluation. This dataset was captured at UC Davis and contains four kinds of QUIC-based services: Google Drive, Youtube, Google Doc, Google Search and Google Music. Its data is captured from several scripts using Selenium WebDriver and AutoIt tools to mimic human behavior.

Before data preprocessing stage, the QUIC dataset has been divided into two parts, one of which is called pretrained dataset with a total of 6,439 flows for unsupervised learning and the other is called retrained dataset with a total of 150 flows for supervised learning. The number of flows contained in each service category in the pretrained dataset and retrained dataset can be seen in Table 1.

Table 1. Flow distribution of QUIC dataset

| Services | Flow size in pretrained dataset | Flow size in retrained dataset |
| --- | --- | --- |
| Google Drive | 1,634 | 30 |
| Youtube | 1,077 | 30 |
| Google Doc | 1,221 | 30 |
| Google Search | 1,915 | 30 |
| Google Music | 592 | 30 |

For the QUIC dataset, we chose to ignore short flows (short flows are those with less than 100 packets before sampling). The number of subflows sampled from short flows is tiny, with insufficient packets in the short flows when sampling.

By using two different flows sampling methods mentioned in 4.1. In the data preprocessing stage, Packets in each flow in the pretrained dataset and retrained dataset are sampled 100 times to form a mass of subflows stored on federated clients and a spot of subflows stored on the central server, respectively. After sampling, there are 528,543 subflows on federated clients and 12,038 subflows on the central server. The sampling result of each service in the pretrained dataset and retrained dataset can be seen in Table 2.

In the federated pretrained stage, all data of the pretrained dataset are used for training on the selected client. In the central server retraining stage, there are 8,711 sampled subflows of the retrained dataset for training and 4,338 for testing on the federated server.

**Table 2.** Sampling result of pretrained dataset and retrained dataset for incremental sampling or simple sampling

| Services | Number of subflows in pretrained dataset | Number of subflows in retrained dataset |
| --- | --- | --- |
| Google Drive | 163,400 | 3,000 |
| Youtube | 107,700 | 3,000 |
| Google Doc | 122,068 | 2,993 |
| Google Search | 76,372 | 112 |
| Google Music | 59,003 | 2,933 |

### 5.3 Sampling Method and Communication Round

In the network classification task, we separately examine the effects of different sampling methods on the federated semi-supervised scenarios we set. In the scenario, we chose to combine federated average aggregation algorithm (FedAvg) [22] with semi-supervision and set up 100 clients and a central server. Before training a deep neural network, the server randomly selects 80% of users from the 100 clients as the training object. After the data preprocessing stage, each client is randomly assigned approximately 5,000 subflows data examples from a pretrained dataset.

As shown in Fig. 5, When the central server communicated with the federated client 100 times, we achieved 97.81% accuracy by incremental sampling and 91.08% accuracy by simple sampling, respectively. Both simple sampling and incremental sampling methods are suitable for sampling streams captured from different parts of the flows. The results show that both can achieve excellent classification accuracy. The accuracy of the incremental sample was about 8% higher than that of the simple sample. When using simple sampling, it is not easy to capture both long and short patterns. Incremental sampling allows sampled flows to contain many packets in short-range and some packets in long-range. That is why incremental sampling outperforms simple sampling methods.

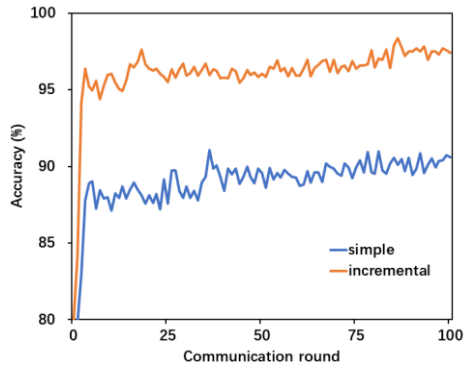

**Fig. 5.** Test accuracy of different sampling methods and communication rounds.

### 5.4 Centralized Semi-supervised vs. FSSL

We also hypothesize a set of comparative experiments. As shown in Table 3, the ultimate stable accuracy of our method is comparable to that of centralized semi-supervised learning. By contrast, the obvious advantage of our method is that it skillfully solves the unavoidable privacy security problem of centralized learning. At the same time, it can also be seen that the superiority of incremental sampling to capture different degrees of patterns is indeed better than that of simple sampling, whether it is centralized training or federal semi-supervised training.

Table 3. Centralized Semi-supervised vs. FSSL

| Sampling method | Centralized | FSSL |
| --- | --- | --- |
| simple | 94.04% | 91.08% |
| incremental | 97.93% | 97.81% |

Meanwhile, we also compared the performance metrics of the two training methods. Compared with the centralized method, the federated semi-supervised method also shows a similar high accuracy score with just a few labeled samples on the QUIC dataset, as shown in Fig. 6.

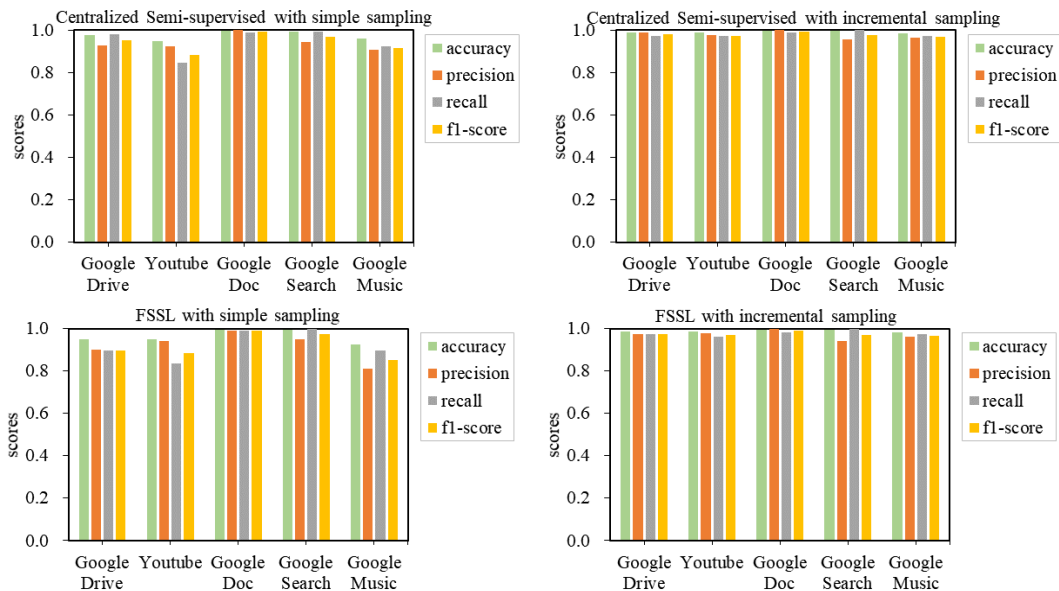

**Fig. 6.** Performance metrics for different network traffic categories with sampling methods on the test dataset: (a) Centralized Semi-supervised with simple sampling (b) Centralized Semi-supervised with incremental sampling (c) FSSL with simple sampling (d) FSSL with incremental sampling

## 6  Conclusion

In this paper, we propose an FSSL approach, which can alleviate the current shortage of labeled datasets and protect user privacy security. We use CNN model, which takes the sampling time series function as input. Five Google services based on the QUIC protocol were used to evaluate our approach. Before training the model, the QUIC data set was preprocessed, and the supervised learning and unsupervised learning data were obtained through simple sampling and incremental sampling. In the federated learning framework, we put large amounts of the unlabeled dataset into a pretrained model to perform unsupervised training on clients. The retraining stage was completed by placing a small set of the labeled dataset in the retrained model on the server. Through the comparison of experiments, it was concluded that (1) Using incremental sampling method conducive to enhance the accuracy of the model; (2) Compared with centralized semi-supervised learning, our method is superior.